\documentclass[draftcls,12pt,onecolumn]{IEEEtran}

\usepackage{cite}
\usepackage[pdftex]{graphicx}
\usepackage[cmex10]{amsmath}
\usepackage{amssymb}
\usepackage{amsfonts}

\usepackage{algorithm}
\usepackage{algorithmic}
\usepackage[caption=false,font=footnotesize]{subfig}
\usepackage{color}
\usepackage{array}
\usepackage{mdwmath}
\usepackage{mdwtab}
\usepackage{eqparbox}
\usepackage[colorlinks=true,urlcolor=black]{hyperref}
\begin{document}

\title{On the Limits of Predictability in Real-World Radio Spectrum State Dynamics: From Entropy Theory to 5G Spectrum Sharing}
\author{Guoru~Ding\IEEEauthorrefmark{1},~Jinlong~Wang\IEEEauthorrefmark{1},~Qihui~Wu\IEEEauthorrefmark{1},~Yu-Dong~Yao\IEEEauthorrefmark{2},\\
Rongpeng Li\IEEEauthorrefmark{3},~Honggang Zhang\IEEEauthorrefmark{3}\IEEEauthorrefmark{4},~and~Yulong Zou\IEEEauthorrefmark{5}

\IEEEauthorblockA{\IEEEauthorrefmark{1}PLA University of Science and Technology}

\IEEEauthorblockA{\IEEEauthorrefmark{2}Stevens Institute of Technology}

\IEEEauthorblockA{\IEEEauthorrefmark{3}Zhejiang University}

\IEEEauthorblockA{\IEEEauthorrefmark{4}Universit$\acute{\rm{e}}$ Europ$\acute{\rm{e}}$enne de Bretagne (UEB) $\&$ Sup$\acute{\rm{e}}$lec}

\IEEEauthorblockA{\IEEEauthorrefmark{5}Nanjing University of Posts and Telecommunications}


}
\maketitle

\begin{abstract}
A range of applications in cognitive radio networks, from adaptive spectrum sensing to predictive spectrum mobility and dynamic spectrum access, depend on our ability to foresee the state evolution of radio spectrum, raising a fundamental question: To what degree is radio spectrum state (RSS) predictable? In this paper, we explore the fundamental limits of predictability in RSS dynamics by studying the RSS evolution patterns in spectrum bands of several popular services, including TV bands, ISM bands, and Cellular bands, etc. From an information theory perspective, we introduce a methodology of using statistical entropy measures and Fano inequality to quantify the degree of predictability underlying real-world spectrum measurements. Despite the apparent randomness, we find a remarkable predictability, as large as $90\%$, in the real-world RSS dynamics over a number of spectrum bands for all popular services. Furthermore, we discuss the potential applications of prediction-based spectrum sharing in 5G wireless communications.
\end{abstract}

\IEEEpeerreviewmaketitle

\section{Introduction}
During the past decades, we have witnessed a dramatic growth in wireless access along with the popularity of smart phones, mobile TVs, and many other wireless services. The ever-increasing demand for high data rates in the face of limited radio spectrum resources has motivated the introduction of cognitive radio (CR), which opens a potential communication paradigm to improve spectrum utilization by allowing secondary users to opportunistically access spectrum holes or white spaces unused by primary users. To enable CR, one fundamental challenge is how to reliably identify when and where spectrum holes exist.

Spectrum sensing and spectrum prediction are known as two effective enabling techniques to identify spectrum holes. Briefly, spectrum sensing determines radio spectrum state (RSS) using various signal detection methods, which has been investigated extensively in the literature (see, e.g., a survey in~\cite{SPMag-sensing-survey}). Complementarily, spectrum prediction infers unknown/unmeasured RSS from historical known/measured spectrum data by exploiting the inherent correlation and/or regularity among them, which has gained increasing attention recently (see, e.g., a survey in \cite{prediction-survey}). Spectrum prediction has many merits, e.g., reducing the sensing time and energy consumption involved in adaptive spectrum sensing~\cite{Time-prediction} and increasing the system throughput via prediction-based dynamic spectrum access~\cite{Yin-TVT}, etc.

To reap these benefits, a number of spectrum prediction techniques have been proposed, such as time series-based prediction, autoregressive model-based prediction, hidden Markov models-based prediction, neural networks-based prediction, and Bayesian inference-based prediction, etc (see, e.g., the survey in \cite{prediction-survey} and the references therein). However, so far it is not clear that for various frequency bands, what the upper-bound performance of various prediction techniques could be. Moreover, the RSS evolution patterns are generally determined by the human's usage of radio spectrum. Although we rarely consider the human activity in radio domain to be totally random, current models of RSS evolution are fundamentally stochastic, see the most widely used continuous/discrete-time Markov chain models in~\cite{Empirical-Modelling}. Yet the probabilistic nature of the existing modeling framework raises fundamental questions: What is the role of randomness in RSS evolution and to what degree is RSS dynamics predictable?

This paper attempts to study the interplay between the regular (and thus predictable) and the random (and thus unforeseeable) underlying real-world RSS dynamics theoretically and provide certain guidance over how to apply the predicted RSS to the design of future wireless communication systems technically. Specifically, from an information theory perspective, we introduce a methodology of using statistical entropy measures and Fano inequality to quantify the degree of predictability underlying real-world spectrum measurements and provide some intuitive thoughts and conclusions. After validating the fundamental limits of predictability in RSS dynamics, this paper moves forwards by addressing the potential applications of the predicted RSS in 5G wireless communications.

\section{Spectrum Data Description and Preprocessing}
In order to ensure the reproducibility of the spectrum prediction analysis in this paper for other researchers, we uses a well-known open source real-world spectrum dataset from the RWTH Aachen University spectrum measurement campaign~\cite{Empirical-Modelling}. In this paper, we are primarily interested in several popular services, including TV bands, ISM bands, and Cellular bands, etc. The resolution bandwidth of each individual spectrum band is 200 kHz. The inter-sample time is about 3 minutes, which corresponds to 3360 samples one week for each 200 kHz spectrum band\footnote{The original inter-sample time in~\cite{Empirical-Modelling} is about 1.8 seconds, which results in 48000 samples one day and 336000 samples one week for each individual spectrum band. To facilitate the presentation and analysis, a preprocessing procedure in this paper is performed to obtain a new spectrum dataset by averaging consecutive 1000 samples.}.

\begin{figure*}[!t]
\centering
\label{fig:DTV}
\includegraphics[width=0.6\linewidth]{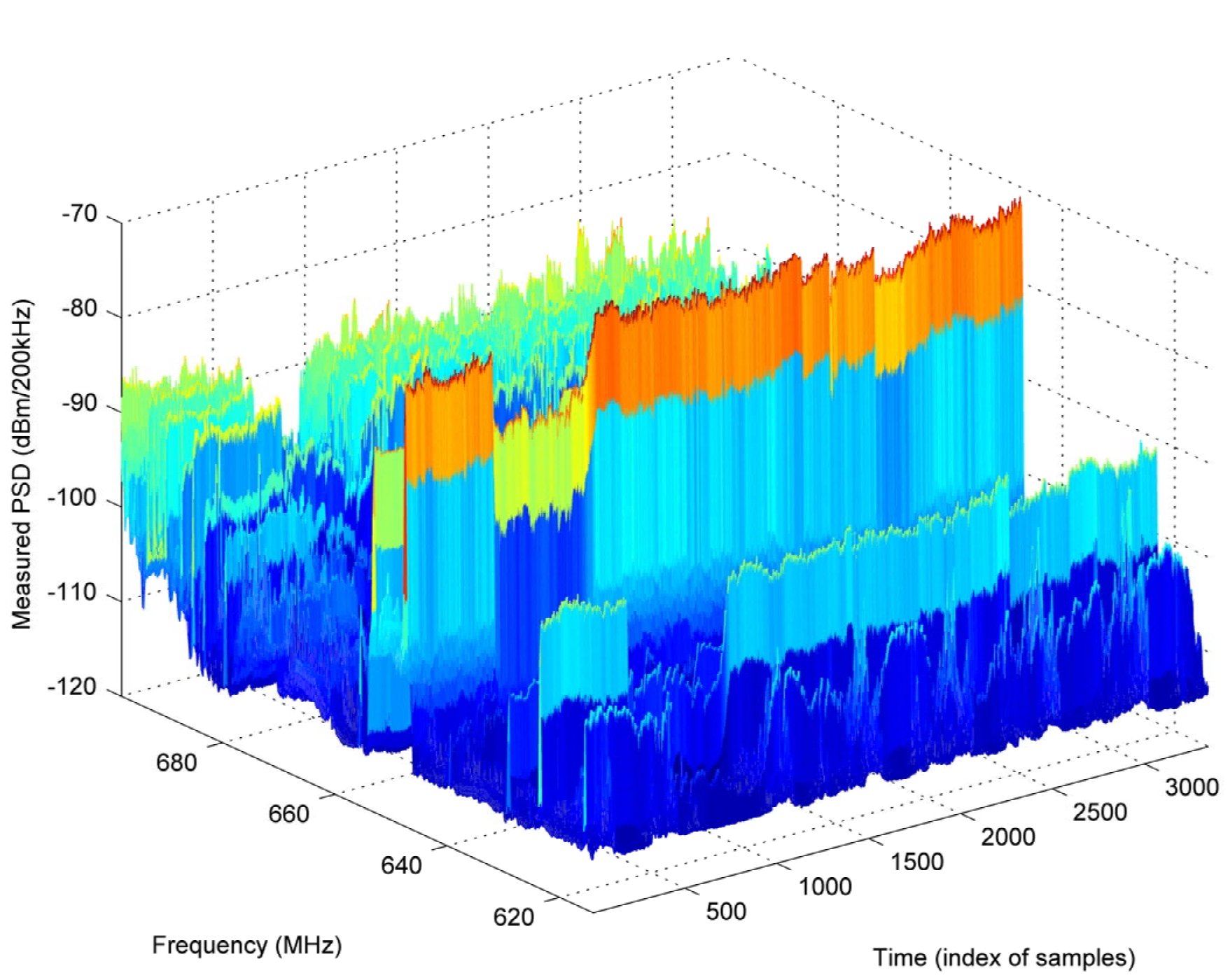}
\caption{The 3-D view of the evolution trajectories of one week real-world RSS in the TV bands ($614\!\sim\!698$ MHz). For each 200 kHz spectrum band, about 3360 samples one week and thus 480 samples one day are plotted.}
\label{fig-RSS-Traces-AB}
\end{figure*}

As an illustrative example, Fig. \ref{fig-RSS-Traces-AB} shows the evolution trajectories of one week real-world RSS, i.e., measured power spectral density (PSD) values, in TV bands. Several interesting phenomena can be observed. First of all, the RSS dynamics for various frequency bands are significantly different, several bands are heavily loaded but others not. Moreover, randomness and regularities coexist in the RSS evolution. Very strong signals can be identified in several TV bands, and, it appears that the temporal variations of signals in these bands are not that significant as those in other bands.

To further show the spectrum utilization of each 200 kHz spectrum band, Fig. \ref{Fig-duty-cycle-across-frequency} plots the duty cycle over the frequency under two well-known detection thresholds. One threshold -107 dBm/200 kHz has initially been proposed in the IEEE 802.22 working group for detection of wireless microphones in 200 kHz channels in the TV bands and the other more sensitive threshold -114 dBm/200 kHz has been specified in the FCC's final rules~\cite{Empirical-Modelling}. As shown in Fig. \ref{Fig-duty-cycle-across-frequency}, the binary spectrum occupancy (BSO) is highly dependent on the selection of the specific detection threshold.

\begin{figure*}[!h]
\centering
\label{fig:DC}
\includegraphics[width=0.5\linewidth]{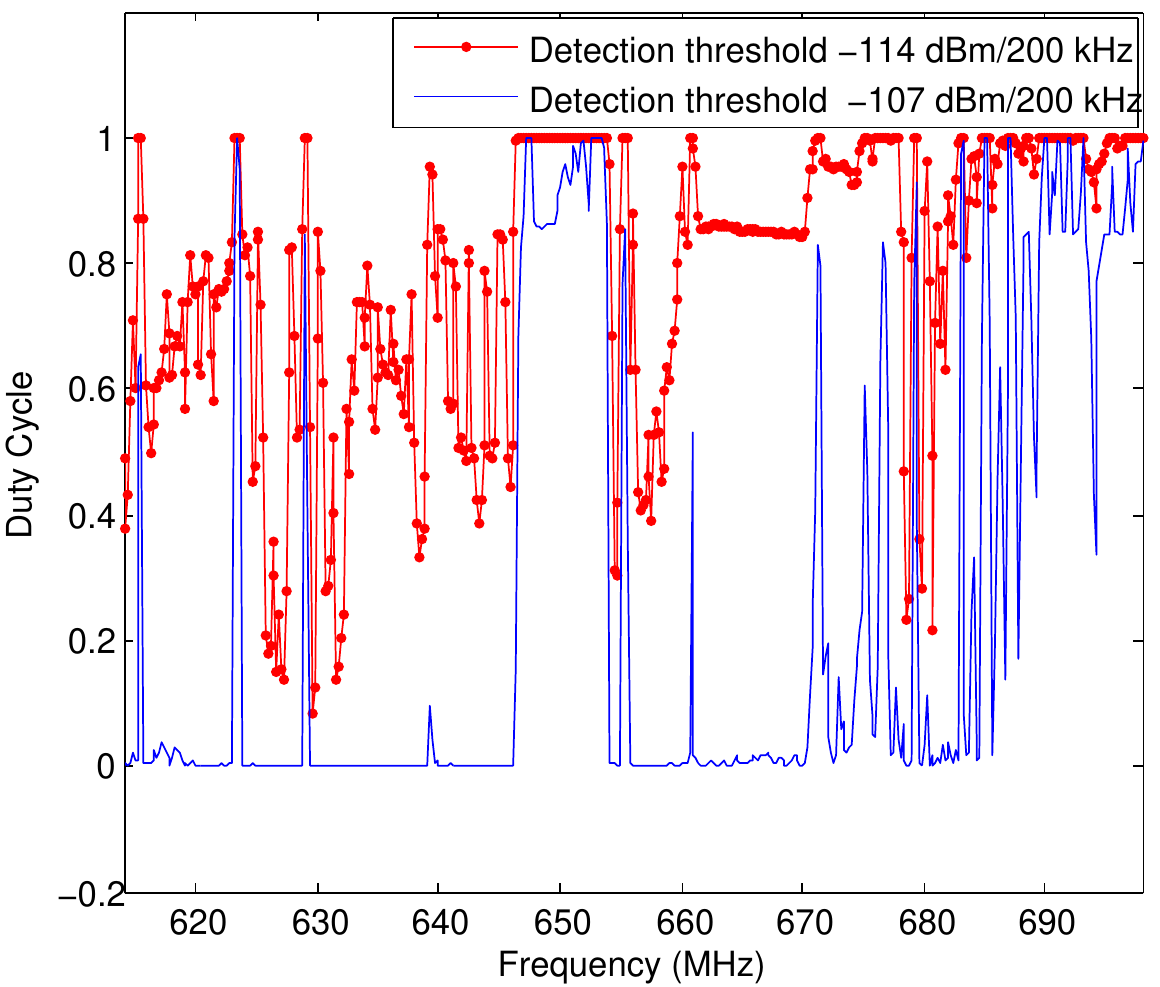}
\caption{Impact of the detection threshold on the duty cycle in TV bands ($614\!\sim\!698$ MHz).}
\label{Fig-duty-cycle-across-frequency}
\end{figure*}

Until now, to characterize the RSS dynamics, most of the existing studies have focused on BSO traces by analyzing the ON and OFF state evolution over time. Instead, in this paper, we will investigate the continuously measured PSD traces and analyze the predictability of the PSD evolution over time, mainly for the following concerns: the PSD is the original raw data, while the BSO, obtained from the PSD by comparing with a detection threshold, inevitably introduces detection or sensing errors (e.g., false alarms and miss detections)~\cite{Haykin_2005}.

\section{Spectrum Prediction Analysis: To What Degree is Radio Spectrum State Predictable?}
In this section, we first perform the prediction analysis on each individual 200 kHz spectrum band separately, and then on the whole spectrum bands allocated to each service statistically.

\subsection{Entropy Analysis}
For a given spectrum band, let $X_i$ be a random variable representing its state at time slot $i$. The state of this band from time slot $1$ to time slot $n$ is a random variable series $X_1, X_2, ..., X_n$. Entropy is probably the most fundamental quantity characterizing the degree of predictability of a random variable series. In general, lower entropy implies higher predictability, and vice versa. Recently, entropy-based analysis have already been introduced in various prediction scenarios such as atmosphere~\cite{atmosphere}, network traffic~\cite{Rongpeng-CMag}, and human mobility~\cite{Limits-of-predictability}. The basic idea is that entropy offers a precise definition of the informational content of predictions and it is renowned for its generality due to minimal assumption on the model of the studied scenario.

Specifically, to facilitate the following entropy analysis of RSS dynamics, we first quantize the PSD values for each individual spectrum band into $Q$ RSS levels. Then, let $S=\{X_1, X_2, ..., X_n\}$ denote the series or sequence of RSS levels occurred at $n$ consecutive time slots and we have the following three entropy measures to characterize the RSS dynamics:
\begin{itemize}
  \item Random entropy $E^{\rm{rand}}= \log_2{Q}$, capturing the degree of predictability of the given spectrum band's evolution if each RSS level occurs with equal probability in each time slot.
  \item Temporal-uncorrelated entropy $E^{\rm{unc}} = -\sum_{i=1}^{n} p_i\log_2p_i,$ where $p_i$ is the probability that the $i$-th RSS level occurred in the sequence $S$. $E^{\rm{unc}}$, also known as Shannon entropy or classical information theoretical entropy, is by far the most often used entropy metric, which characterizes the heterogeneity of the RSS evolution patterns without taking into account the history of the process.
  \item Actual entropy $E^{\rm{actual}}=-\sum_{S_i\subset S}P(S_i)\log_2P(S_i)$, where $P(S_i)$ is the probability of a particular time-ordered subsequence $S_i$ occurred in the trajectory of $S$. Thus, $E^{\rm{actual}}$ depends not only on the occurrence frequency of each RSS level, but also the temporal order in which the RSS levels occurred, and it captures the full frequency-time structure present in a given spectrum band's revolution pattern. In practice, to calculate the actual entropy from the historical spectrum measurements, we use an estimator based on Lempel-Ziv data compression~\cite{Kontoyiannis}, which is known to rapidly converge to the actual entropy of a time series. For a time series with length $n$, the entropy is estimated by $E^{\rm{actual}}_{\rm{est}} = (\frac{1}{n}\sum_{i=1}^n \Lambda_i)^{-1}\ln n,$ where $\Lambda_i$ is the length of the shortest subsequence starting at the $i$-th time slot which doesn't previously appear from time slot 1 to time slot $i$.
\end{itemize}

\begin{figure*}[!t]
\centering
\label{fig:DC}
\includegraphics[width=0.5\linewidth]{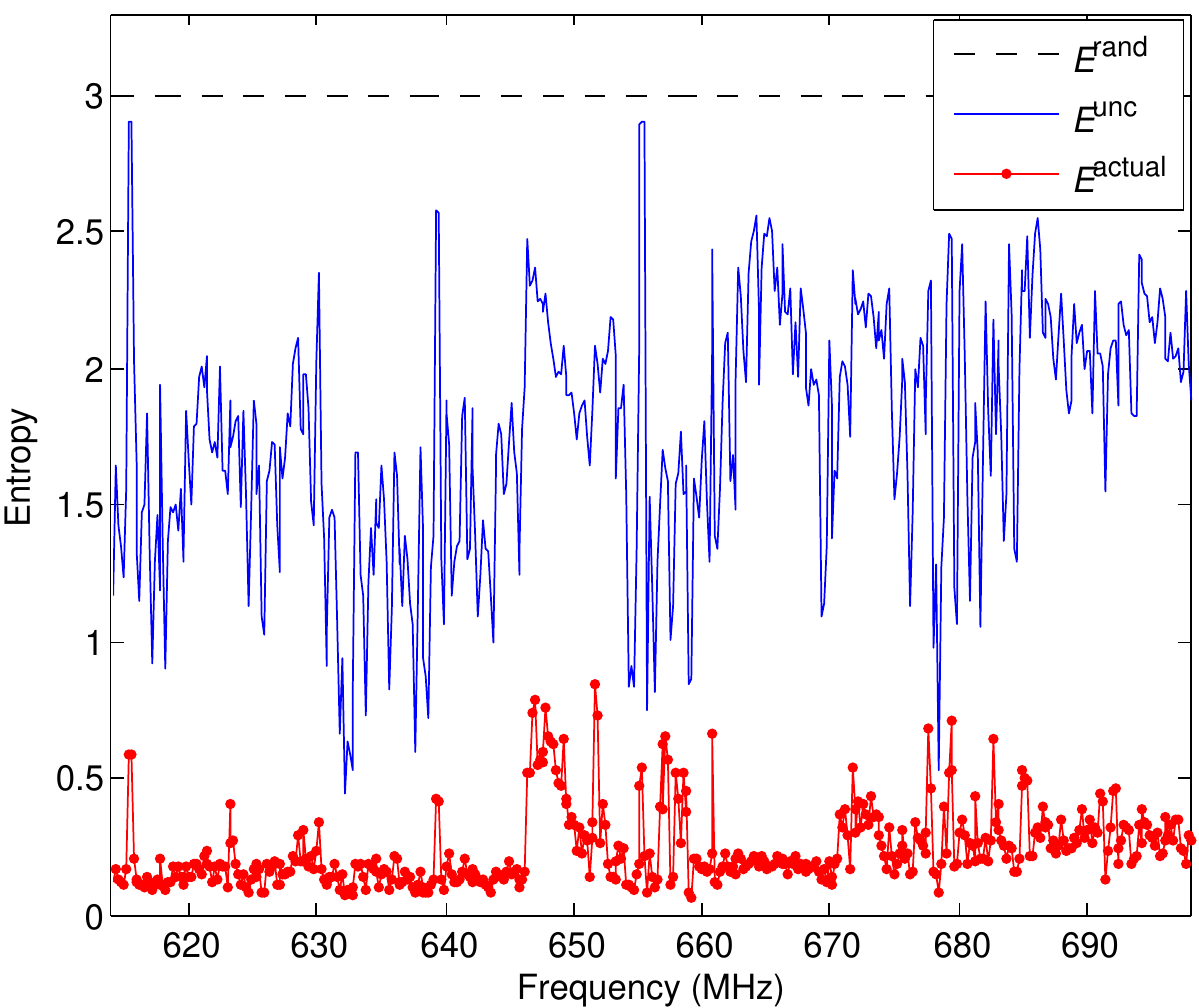}
\caption{The entropy of the RSS dynamics in TV bands. The number of RSS levels is set as $Q=8$ for each individual 200 kHz spectrum band and thus the random entropy $E^{\rm{rand}}= \log_2{(Q=8)}=3$ bits.}
\label{Fig-Entropy-across-frequency}
\end{figure*}

Intuitively, we have $0 \le E^{\rm{actual}} \le E^{\rm{unc}} \le E^{\rm{rand}}$, which is illustrated in Fig. \ref{Fig-Entropy-across-frequency} via analyzing the real-world spectrum measurements in TV bands. Extremely, if a spectrum band has actual entropy $E^{\rm{actual}}=0$, its RSS evolution is completely regular and thus fully predictable. If, however, a spectrum band's actual entropy $E^{\rm{actual}}=E^{\rm{rand}}=\log_2Q$, its trajectory is expected to follow a quite random pattern and thus we cannot predict it with an accuracy exceeding $1/Q$. As shown in Fig. \ref{Fig-Entropy-across-frequency}, all spectrum bands have finite actual entropies between 0 and $E^{\rm{rand}}$, indicating that not only a certain amount of \emph{randomness} governs their future whereabouts, but also that there is some \emph{regularity} in their dynamics that can be exploited for predictive purposes.

Based on the obtained actual entropy, in the following, we aim to quantify the limits of the predictability of a spectrum band's next state based on its trajectory history. That is, we want to answer the question: How predictable is a spectrum band's next state given the entropy of its historical trajectory?

\subsection{Predictability Analysis}

An important measure of predictability is the probability $\it{\Pi}$ that an appropriate predictive algorithm can correctly predict a spectrum band's future state. This quantity is subject to Fano's inequality~\cite{Fano-inequality}. That is, if an individual spectrum band with an actual entropy $E^{\rm{actual}}$ evolutes between $Q$ RSS levels, its predictability ${\it{\Pi}} \le {\it{\Pi}}^{\max}$, where $\it{\Pi}^{\max}$ is determined by

$E^{\rm{actual}} = -[{\it{\Pi}}^{\rm{max}} \log_2 {\it{\Pi}}^{\rm{max}}+ (1-{\it{\Pi}}^{\rm{max}})\log_2 (1-{\it{\Pi}}^{\rm{max}})]+(1-{\it{\Pi}}^{\rm{max}})\log_2 (Q-1).$

Based on this relationship, for each spectrum band, we can obtain the upper-bound predictability, ${\it{\Pi}}^{\rm{max}}$, through numerical calculations given $Q$ and $E^{\rm{actual}}$.

As an illustrative example, Fig. \ref{Fig-predictability-across-frequency} shows the upper-bound predictability ${\it{\Pi}}^{\rm{max}}$ over each 200 kHz TV band separately when the number of RSS levels is set as $Q=8.$ For comparison, the predictability of independent identical distributed (i.i.d.) Gaussian noise data with one-week samples is also plotted. We have the following observations:
\begin{itemize}
  \item The predictability of real-world RSS data varies significantly for different spectrum bands. For example, there are a number of TV bands with the predictability higher than 0.95, which means that at most $5\%$ of the time these spectrum bands change their states in a manner that appears to be random, and in the remaining $95\%$ of the time we can expect to predict their whereabouts. On the other hand, we also see that there are a few TV bands with the predictability lower than 0.9, which means that no matter how good our predictive algorithms, we cannot predict with better than $90\%$ accuracy the future states of these spectrum bands.
  \item For all TV bands, the predictability of real-world RSS data are much higher than that of the i.i.d Gaussian noise data. This demonstrates that the temporal correlation or regularity in the real-world RSS data benefits the predictability.
\end{itemize}

\begin{figure*}[!t]
\centering
\label{fig:DC}
\includegraphics[width=0.5\linewidth]{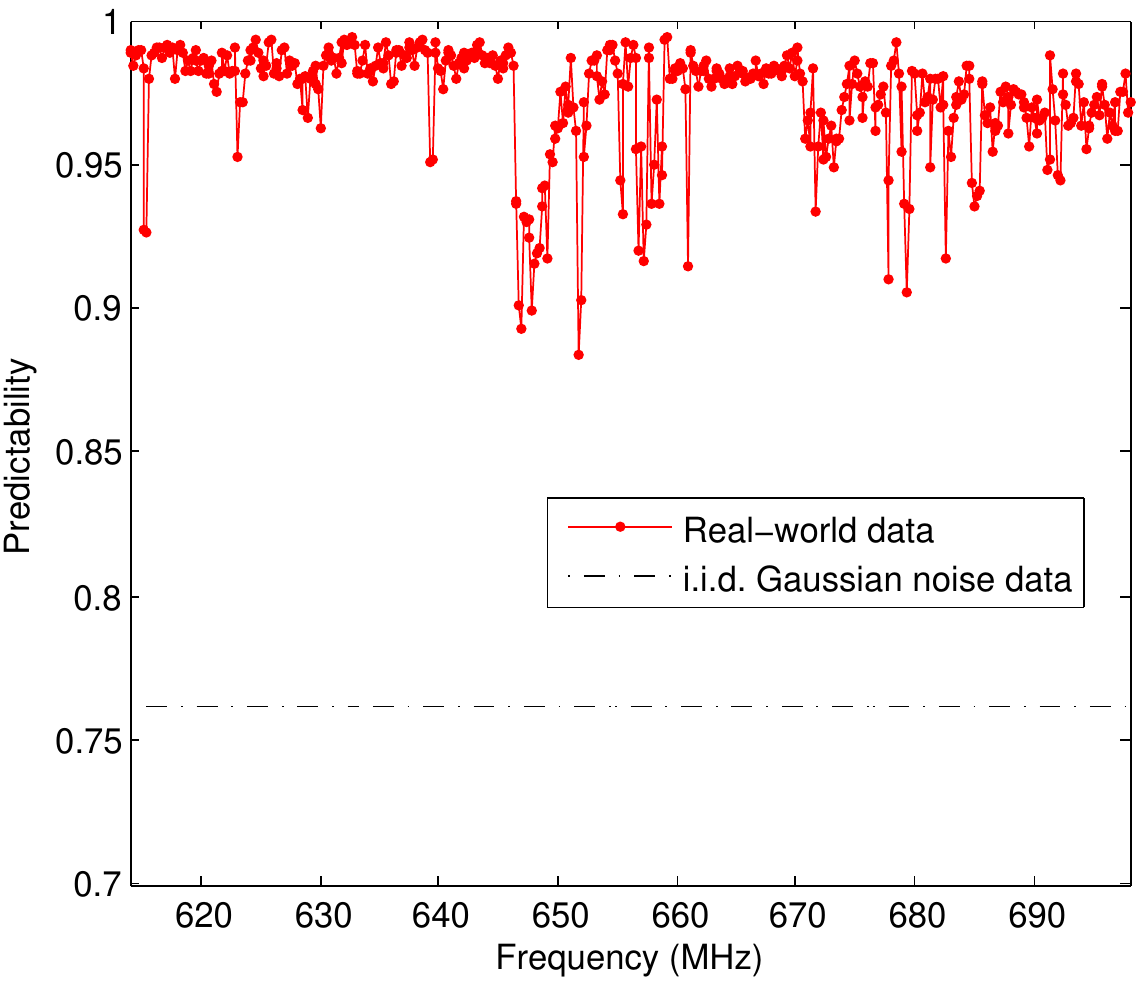}
\caption{The predictability in RSS dynamics for TV bands.}
\label{Fig-predictability-across-frequency}
\end{figure*}

Furthermore, from a statistical perspective, Fig. \ref{Fig-CDF-predictability} shows the cumulative distribution functions (CDFs) for the predictability (with $Q=8$) of various services, including TV bands ($614\!\sim\!698$ MHz), ISM bands ($2400.1\!\sim\!2483.3$ MHz), cellular bands (GSM1800 uplink $1710.2\!\sim\!1784.8$ MHz and GSM1800 downlink $1820.2\!\sim\!1875.4$ MHz), and 2.3 GHz bands ($2300\!\sim\!2400$ MHz)\footnote{The predictability results on 2.3 GHz bands are include in Fig. 5, since Europe is currently looking at deploying licensed/authorized shared access within these bands.}. We have the following observations:
\begin{itemize}
  \item Among all services, TV bands have the steepest CDF, with the minimum predictability 0.8836. Comparatively, most bands in 2.3 GHz have relatively low predictability, with the minimum close to the predictability of i.i.d. Gaussian noise data, 0.7623. ISM bands have a CDF between TV bands and cellular bands, which implies that a larger (lower) proportion of ISM bands have higher predictability than those of cellular (TV) bands.
  \item A predictability superiority of the GSM1800 downlink is observed over the GSM1800 uplink for spectrum bands with predictability levels in the bottom 70 percent. However, for spectrum bands with predictability levels in the top 30 percent, a predictability superiority of the GSM1800 uplink is observed over the GSM1800 downlink. That is, although majority of the GSM1800 downlink bands have superior predictability, there are some GSM1800 uplink bands have very high predictability levels. This somewhat conflicting observations might result from the fact that quite regular patterns of human¡¯s spectrum usage exist in few GSM1800 uplink bands.
\end{itemize}

\begin{figure*}[!t]
\centering
\label{fig:stacksub:a}
\includegraphics[width=0.5\linewidth]{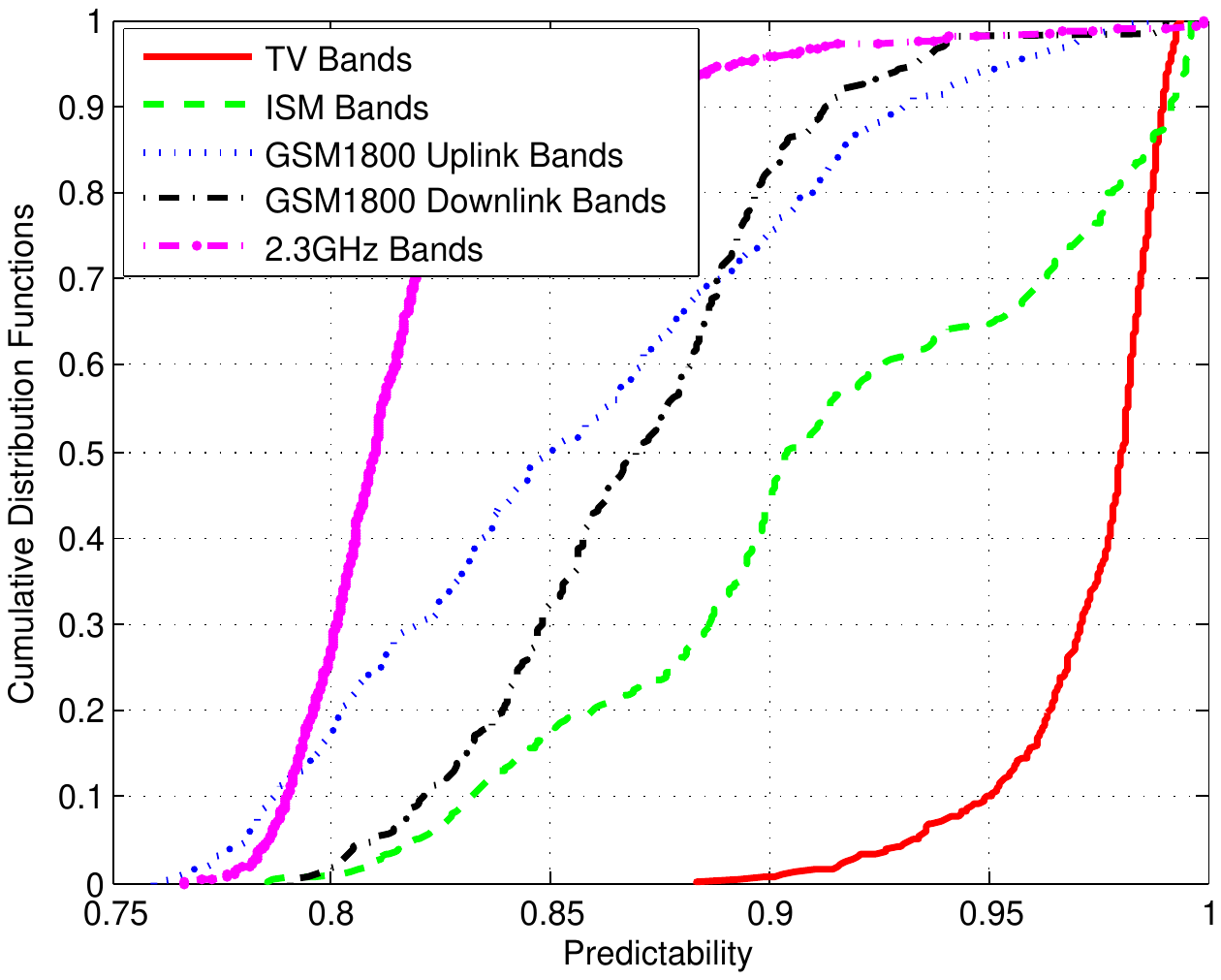}
\caption{The cumulative distribution functions (CDFs) for the predictability of various services.}
\label{Fig-CDF-predictability}
\end{figure*}

\section{Applications: 5G Spectrum Sharing}
Radio spectrum usage is an essential issue in 5G wireless communications~\cite{JGA-5G}. The explosion of data rates offered by mobile internet and internet of things (IoTs) is overwhelming allocated 2G/3G/4G radio spectrum. In the past, new cellular spectrum has typically been made available through spectrum refarming. However, clearing radio spectrum from an allocated but under-utilized usage to repurpose the spectrum band to another usage often requires many years to accomplish, which makes it difficult to keep pace with user demand of gigabit per second (Gbps) data rates for 5G~\cite{Mitola-5G}. On the other hand, technological innovations such as millimeter wave communications and visible light communications can offer very high data rates; However, these disruptive technologies are mainly for small cells and low mobility usage. To provide wide area cell types, spectrum resources below 3 GHz will be needed~\cite{CACB-2014}.

\begin{figure*}[!t]
\centering
\label{fig:stacksub:a}
\includegraphics[width=0.6\linewidth]{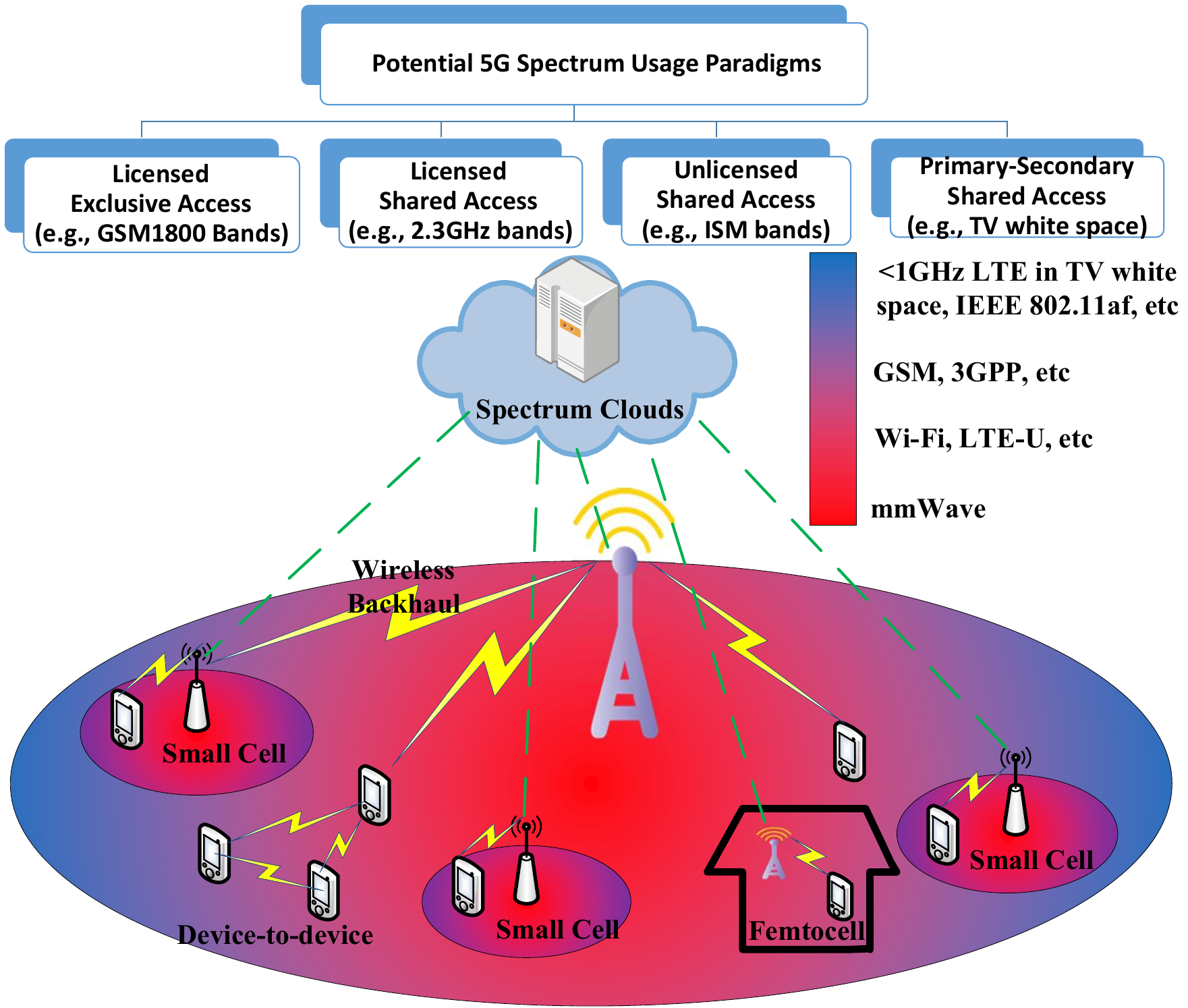}
\caption{A vision for potential 5G spectrum usage paradigms.}
\label{Fig-5G-Spectrum-Usage}
\end{figure*}

To address these challenges, spectrum sharing is contemplated as the primary candidate, which has been well recognized as an affordable, near-term solution of meeting the 5G radio spectrum requirements and increasing radio access network capacities for 5G content delivery. Specifically, \emph{5G spectrum sharing is well beyond the previous studies on cognitive radio-based spectrum sharing}, since one main feature of the latter is the opportunistic primary-secondary access in unlicensed bands (such as TV white space). In contrast, as shown in Fig. \ref{Fig-5G-Spectrum-Usage}, 5G spectrum sharing may occur in both licensed bands (e.g., GSM1800 bands, 2.3 GHz bands) and unlicensed bands (e.g., ISM bands, TV bands). Moreover, one distinguished feature of potential 5G spectrum usage is the \emph{diversity}, i.e., besides the licensed exclusive access in traditional cellular networks, licensed/authorized shared access, unlicensed shared access (known as LTE in unlicensed bands), and primary-secondary access will coexist~\cite{White-Paper-5G}.

Spectrum prediction will play a significant role in 5G spectrum sharing. Several potential applications are described below:
\begin{itemize}
  \item \emph{Cost-efficient wideband carrier aggregation}. To meet the 5G capacity requirement, it is known that no single band or air interface standard by itself fully suffices, and, it is inevitable for 5G devices to aggregate the benefits of multiple (non-continuous) spectrum bands of a very wide range, possibly, from several hundreds MHz bands to 30-300 GHz millimeter wave bands. Consequently, proactive schemes are expected to exploit the evolution dynamics of various spectrum bands of such a wide range, and enable wideband carrier selection and aggregation in a timely and cost-efficient manner.
  \item \emph{Dynamic frequency selection and predictive interference mitigation}. One dominant theme for wireless evolution into 5G is network densification, which is realized mainly by increasing the density of infrastructure nodes (such as base stations and relays) in a given geographic area. It is anticipated that hyper-dense small cells are largely privately owned, and of unplanned deployment. The small cells thus need to be capable of being configured, optimized and healed by themselves to select the communication frequency bands and not to cause any noticeable interference to the existing neighborhood networks. The knowledge from spectrum prediction can be used by the small cells to assist such autonomous processes through dynamic frequency selection and predictive interference mitigation.
\end{itemize}

\section{Conclusion and Discussions}
Predicting the radio spectrum state evolution gains increasingly attention as the explosive growing demand for dynamic spectrum access. In this paper, statistical entropy measures and Fano inequality are exploited to quantify the degree of predictability underlying real-world spectrum measurements. The results in this paper, serving as the upper-bound prediction performance, can provide a performance bound of various predictive algorithms and a general guidance to the design of future wireless communication systems. Notably, it remains a challenge for the state-of-the-art prediction techniques to obtain a prediction precision approaching to the upper-bound predictability. Further improvement of the forecast accuracy of spectrum prediction techniques in a real-time mode are thus required.

\section*{Acknowledgment}
We gratefully acknowledge the use of wireless data from Spectrum Data Archive of the Institute of Networked Systems at RWTH Aachen University. We sincerely appreciate the helpful comments and suggestions from the Guest Editor Prof. Mischa Dohler and the anonymous reviewers, which have helped us improve this article significantly. We also thank Mr. Youming Sun for his helpful input in preparing Fig. \ref{Fig-5G-Spectrum-Usage}. This work is supported by the National Natural Science Foundation of China (Grant No. 61301160 and No. 61172062).

\newpage

~~~~~~~~~~~~~~~~~~~~~~~~~~~~~~~~~~~~~~~~~~~~~~~BIOGRAPHIES

GUORU DING (dingguoru@gmail.com) received his B.S. degree in electrical engineering from Xidian University, Xi'an, China, in 2008 and his Ph.D. degree in communications and information systems in College of Communications Engineering, PLA University of Science and Technology, Nanjing, China, in 2014. His research interests include wireless security, cognitive radio networks, machine learning, and big data analytics over wireless networks. He was a recipient of the Best Paper Awards from IEEE VTC 2014-Fall and IEEE WCSP 2009. He actively participates in international standardization association IEEE DySPAN Standards Committee and acts as the Secretary of IEEE 1900.6 and one of the voting members both in IEEE 1900.7 and IEEE 1900.6.

JINLONG WANG (wjl543@sina.com) received his B.S. degree in wireless communications, M.S. degree and Ph.D. degree in communications and electronic systems from Institute of Communications Engineering, Nanjing, China, in 1983, 1986 and 1992, respectively. He is currently a Chair Professor at PLA University of Science and Technology, Nanjing, China. He is also the co-chair of IEEE Nanjing Section. He has published widely in the areas of signal processing for wireless communications and networking. His current research interests include soft defined radio, cognitive radio, and green wireless communication systems.

QIHUI WU (wqhtxdk@aliyun.com) received his Ph.D. degree in communications and information systems from Institute of Communications Engineering, Nanjing, China, in 2000. From 2003 to 2005, he was a Postdoctoral Research Associate at Southeast University, Nanjing, China. From 2005 to 2007, he was an Associate Professor with PLA University of Science and Technology, Nanjing, China, where he is currently a Full Professor. From March 2011 to September 2011, he was an Advanced Visiting Scholar in Stevens Institute of Technology, Hoboken, USA. His research interests include software defined radio, cognitive radio, and smart radio.

YU-DONG YAO (yyao@stevens.edu) has been with Stevens Institute of Technology, Hoboken, New Jersey, since 2000, and is currently a professor and department director of electrical and computer engineering. Previously, from 1989 and 1990, he was at Carleton University, Ottawa, Canada, as a research associate working on mobile radio communications. From 1990 to 1994, he was with Spar Aerospace Ltd., Montreal, Canada, where he was involved in research on satellite communications. From 1994 to 2000, he was with Qualcomm Inc., San Diego, California, where he participated in research and development in wireless CDMA systems.

RONGPENG LI (lirongpeng@zju.edu.cn) is pursuing his Doctorate degree in the Department of Information Science and Electronic Engineering, Zhejiang University, China. From September to December 2013 he was a visiting doctoral student in Sup¨¦lec, France. From 2006 to 2010 he studied in the Honor Class, School of Telecommunications Engineering, Xidian University, China, and received his B.E. as an ``Outstanding Graduate'' in June, 2010. His research interests focus on Green Cellular Networks, Applications of Reinforcement Learning, and Analysis of Cellular Network Data.

HONGGANG ZHANG (honggangzhang@zju.edu.cn) is a Professor at Zhejiang University, China, as well as the International Chair Professor of Excellence for Universit$\acute{\rm{e}}$ Europ$\acute{\rm{e}}$enne de Bretagne (UEB) $\&$ Sup$\acute{\rm{e}}$lec, France. He is also an Honorary Visiting Professor at the University of York, UK. He served as the Chair of the Technical Committee on Cognitive Networks of the IEEE Communications Society from 2011¨C2012. He is currently involved in research on Green Communications and was the Lead Guest Editor of the IEEE Communications Magazine special issues on ``Green Communications.''

YULONG ZOU (yulong.zou@njupt.edu.cn) is a Full Professor and Doctoral Supervisor at the Nanjing University of Posts and Telecommunications (NUPT), Nanjing, China. He received the B.Eng. degree in Information Engineering from NUPT, Nanjing, China, in July 2006, the first Ph.D. degree in Electrical Engineering from the Stevens Institute of Technology, New Jersey, the United States, in May 2012, and the second Ph.D. degree in Signal and Information Processing from NUPT, Nanjing, China, in July 2012. His research interests span a wide range of topics in wireless communications and signal processing, including the cooperative communications, cognitive radio, wireless security, and energy-efficient communications. Dr. Zou is currently serving as an editor for the IEEE Communications Surveys \& Tutorials, IEEE Communications Letters, EURASIP Journal on Advances in Signal Processing, and KSII Transactions on Internet and Information Systems, etc.

\newpage
\begin{figure*}[!t]
\centering
\label{fig:DTV}
\includegraphics[width=0.6\linewidth]{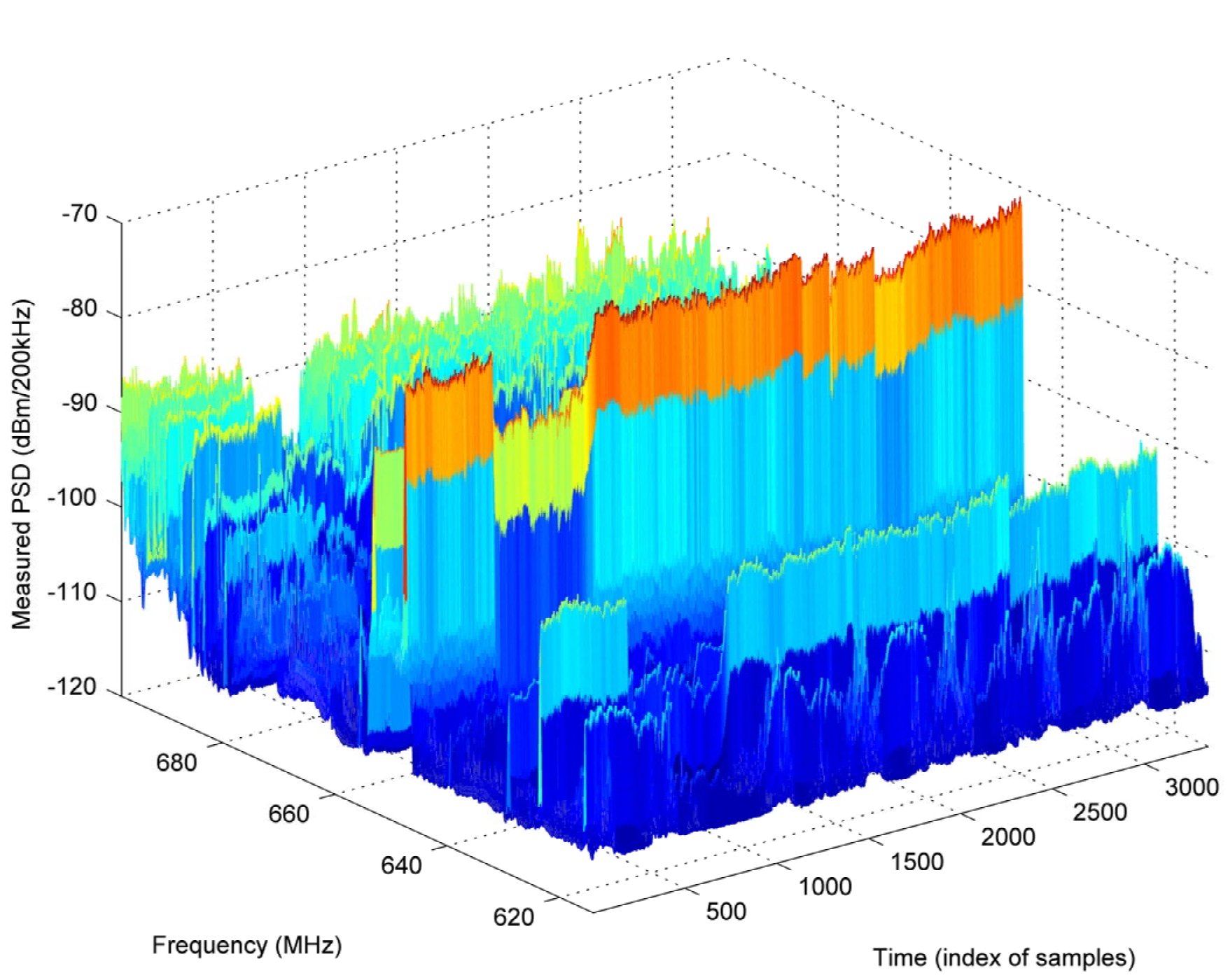}
\caption{The 3-D view of the evolution trajectories of one week real-world RSS in the TV bands ($614\!\sim\!698$ MHz). For each 200 kHz spectrum band, about 3360 samples one week and thus 480 samples one day are plotted.}
\label{fig-RSS-Traces-AB}
\end{figure*}

\newpage
\begin{figure*}[!h]
\centering
\label{fig:DC}
\includegraphics[width=0.5\linewidth]{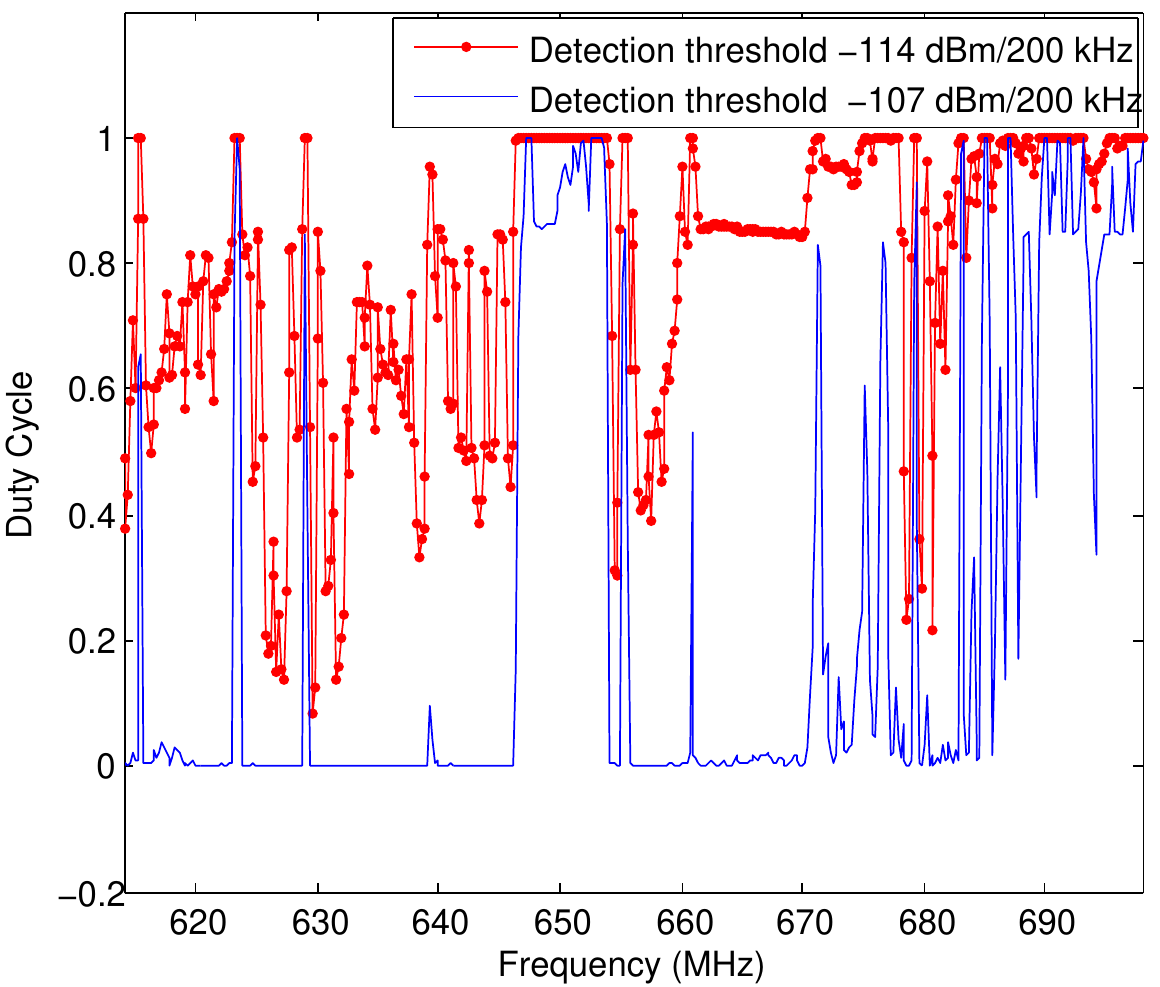}
\caption{Impact of the detection threshold on the duty cycle in TV bands ($614\!\sim\!698$ MHz).}
\label{Fig-duty-cycle-across-frequency}
\end{figure*}

\newpage
\begin{figure*}[!t]
\centering
\label{fig:DC}
\includegraphics[width=0.5\linewidth]{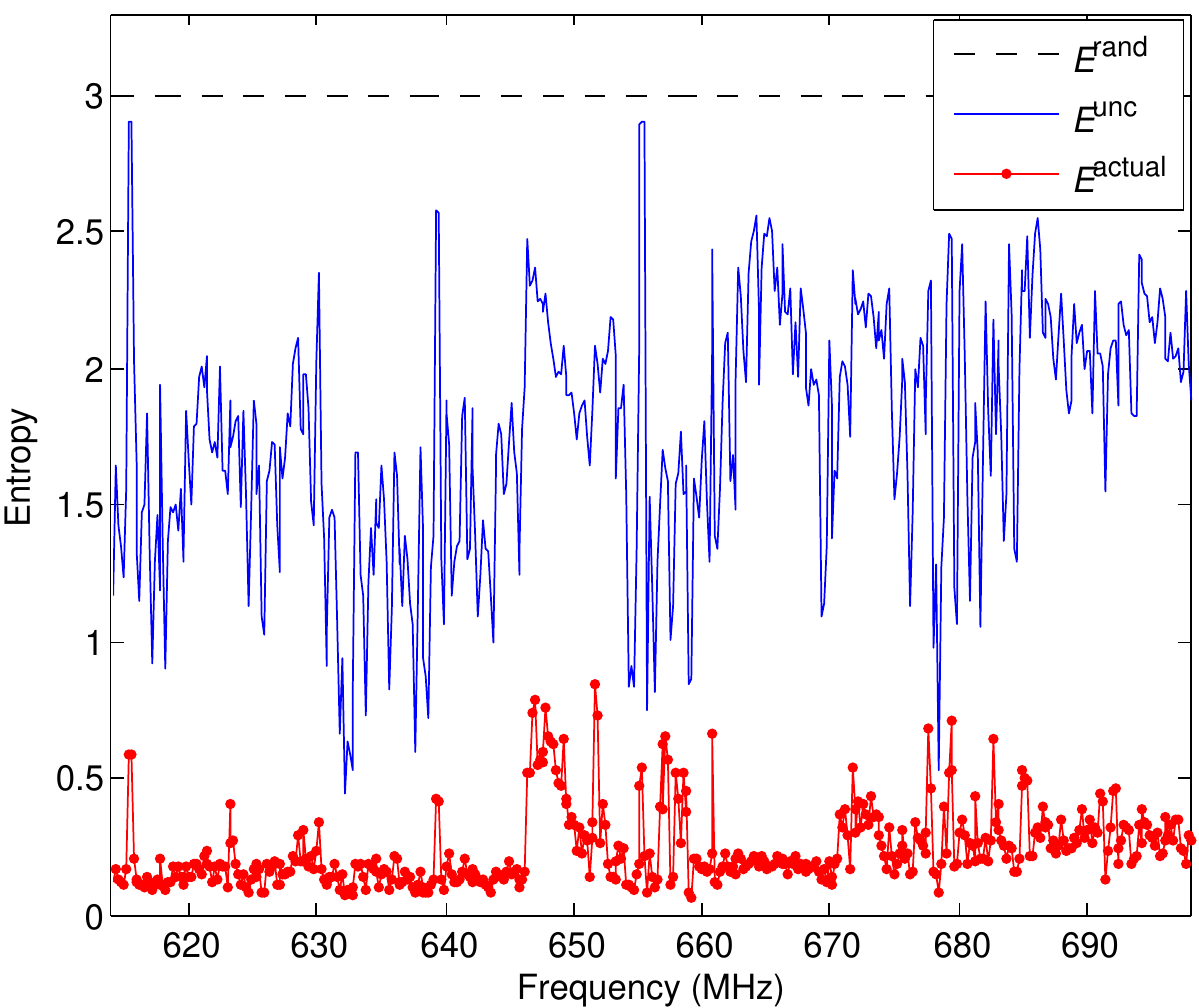}
\caption{The entropy of the RSS dynamics in TV bands. The number of RSS levels is set as $Q=8$ for each individual 200 kHz spectrum band and thus the random entropy $E^{\rm{rand}}= \log_2{(Q=8)}=3$ bits.}
\label{Fig-Entropy-across-frequency}
\end{figure*}

\newpage
\begin{figure*}[!t]
\centering
\label{fig:DC}
\includegraphics[width=0.5\linewidth]{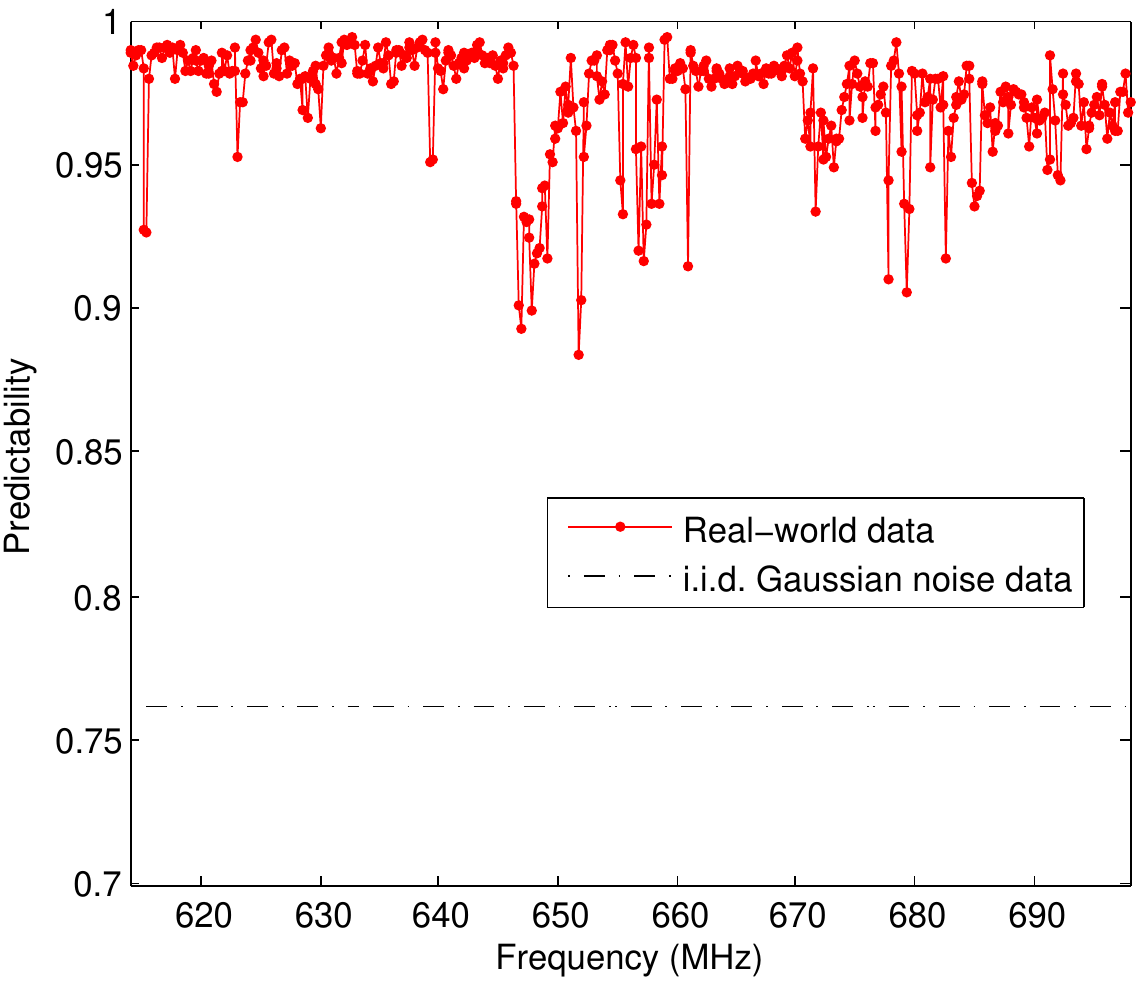}
\caption{The predictability in RSS dynamics for TV bands.}
\label{Fig-predictability-across-frequency}
\end{figure*}

\newpage
\begin{figure*}[!t]
\centering
\label{fig:stacksub:a}
\includegraphics[width=0.5\linewidth]{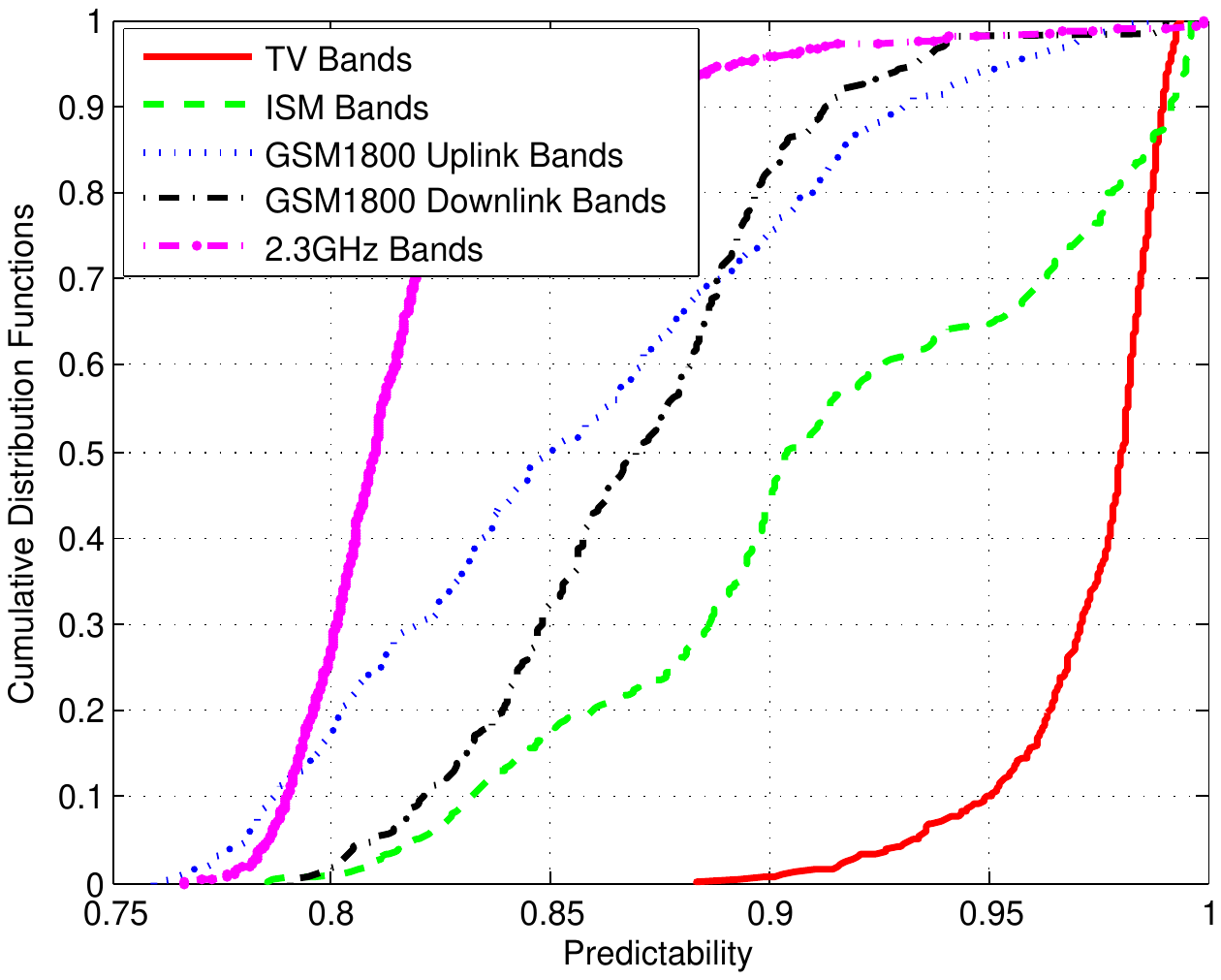}
\caption{The cumulative distribution functions (CDFs) for the predictability of various services.}
\label{Fig-CDF-predictability}
\end{figure*}

\newpage
\begin{figure*}[!t]
\centering
\label{fig:stacksub:a}
\includegraphics[width=0.6\linewidth]{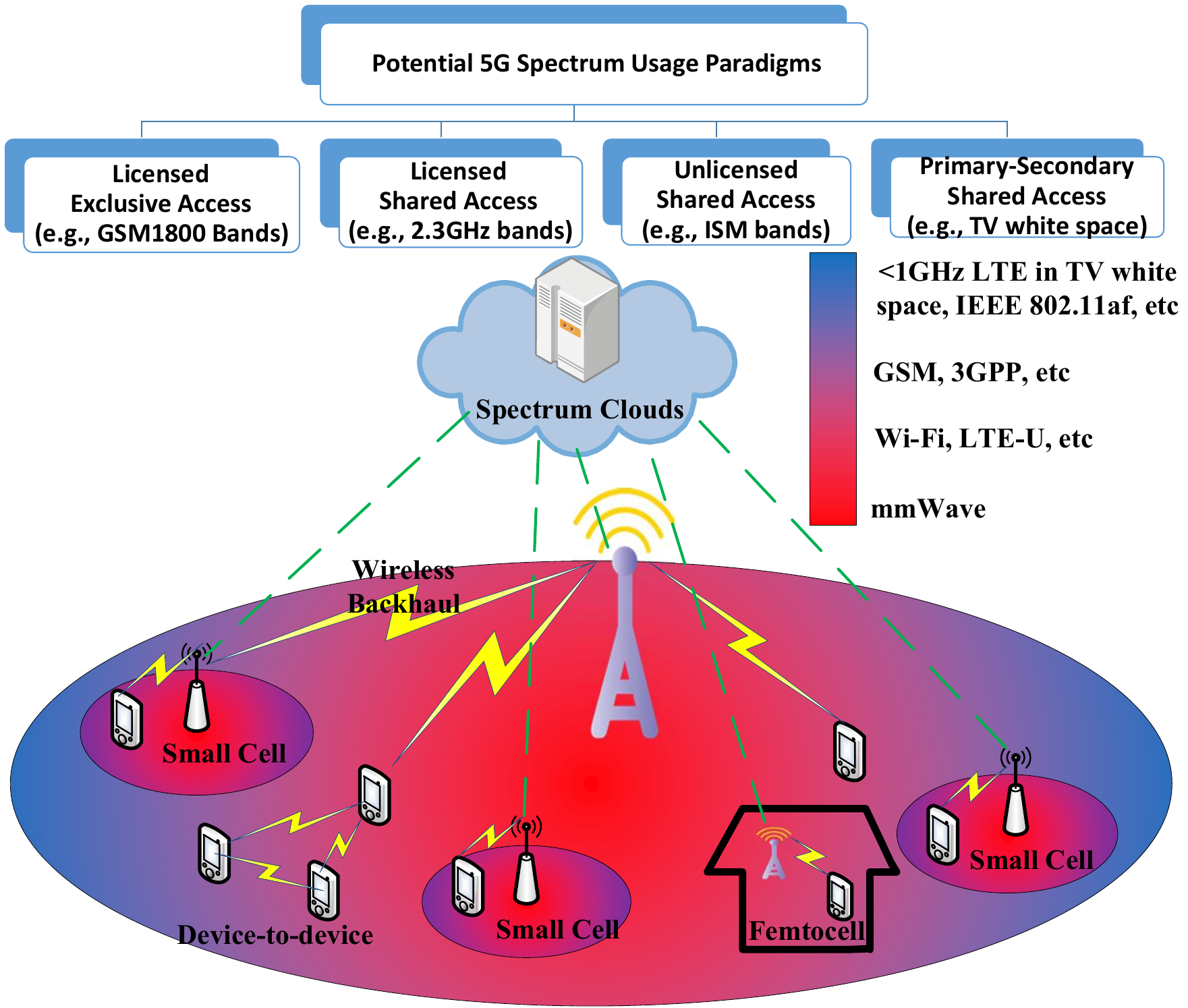}
\caption{A vision for potential 5G spectrum usage paradigms.}
\label{Fig-5G-Spectrum-Usage}
\end{figure*}

\end{document}